\documentclass[iop, numberedappendix]{mn2e}

\usepackage[pdftex]{graphicx}

\usepackage{amsmath, amsthm, amssymb}
\let\bibhang\relax
\usepackage{natbib}
\usepackage{subfigure}
\usepackage{upgreek}
\usepackage{accents}
\usepackage{color}
\usepackage{hyperref}
\usepackage{dblfloatfix}
\usepackage[T1]{fontenc}
\usepackage{aecompl}
\usepackage{enumitem}
\newcommand{\angstrom}{\text{\normalfont\AA}}

\hypersetup{
  colorlinks   = true, 
  urlcolor     = blue, 
  linkcolor    = blue, 
  citecolor   = blue 
}

\newcommand{\Z}{\mathcal{Z}}

\newcommand{\Omnow}{\Omega_{\text{m},0}}
\newcommand{\Obnow}{\Omega_{\text{b},0}}

\newcommand{\Tcmb}{T_{\gamma}}

\newcommand{\Ts}{T_{\text{S}}}
\newcommand{\Tk}{T_{\text{K}}}

\newcommand{\zrei}{z_{\text{rei}}}

\newcommand{\Tvir}{T_{\text{vir}}}

\newcommand{\lsfrunits}{\text{erg} \ \text{s}^{-1} \ \text{Hz}^{-1} \ (M_{\odot} \ \text{yr}^{-1})^{-1}}
\newcommand{\sfrunits}{M_{\odot} \ \text{yr}^{-1}}
\newcommand{\LUV}{L_{\text{UV}}}
\newcommand{\kappaUV}{\kappa_{\text{UV}}}

\newcommand{\nH}{n_{\text{H}}}

\newcommand{\HII}{\text{H} {\textsc{ii}}}

\newcommand{\sfr}{\dot{M}_{\ast}}
\newcommand{\sfrd}{\dot{\rho}_{\ast}}

\newcommand{\Mpeak}{M_{\text{peak}}}
\newcommand{\alphalo}{\alpha_{\mathrm{lo}}}
\newcommand{\alphahi}{\alpha_{\mathrm{hi}}}

\newcommand{\QHII}{Q_{\textsc{HII}}}
\newcommand{\Mmin}{M_{\min}}
\newcommand{\Tmin}{T_{\min}}

\newcommand{\xHII}{x_{\text{H } \textsc{ii}}}

\newcommand{\fstar}{f_{\ast}}

\newcommand{\fcoll}{f_{\text{coll}}}

\newcommand{\Nion}{N_{\text{ion}}}
\newcommand{\Nlw}{N_{\text{LW}}}
\newcommand{\fesc}{f_{\text{esc}}}

\newcommand{\MUV}{M_{\text{UV}}}

\newcommand{\dTb}{\delta T_b}

\title[Cosmic dawn modeling systematics]{Systematic uncertainties in models of the cosmic dawn}
\author[Mirocha, Lamarre, \& Liu]{Jordan Mirocha,\textsuperscript{\thanks{jordan.mirocha@mcgill.ca}}\textsuperscript{\thanks{CITA National Fellow}} Henri Lamarre,
Adrian Liu\textsuperscript{\thanks{acliu@physics.mcgill.ca}} \\
Department of Physics \& McGill Space Institute, McGill University, 3600 Rue University, Montr\'eal, QC, H3A 2T8 \\
}

\begin{document}

\pagerange{\pageref{firstpage}--\pageref{lastpage}} \pubyear{2020}
\maketitle

\begin{abstract}
Models of the reionization and reheating of the intergalactic medium (IGM) at redshifts $z \gtrsim 6$ continue to grow more sophisticated in anticipation of near-future 21-cm, cosmic microwave background, and galaxy survey measurements. However, there are many potential sources of systematic uncertainty in models that could bias and/or degrade upcoming constraints if left unaccounted for. In this work, we examine three commonly-ignored sources of uncertainty in models for the mean reionization and thermal histories of the IGM: the underlying cosmology, halo mass function (HMF), and choice of stellar population synthesis (SPS) model. We find that cosmological uncertainties affect the Thomson scattering optical depth at the few percent level and the amplitude of the global 21-cm signal at the $\sim$5-10 mK level. The differences brought about by choice of HMF and SPS models are more dramatic, comparable to the $1 \sigma$ error-bar on $\tau_e$ and a $\sim 20$ mK effect on the global 21-cm signal amplitude. Finally, we jointly fit galaxy luminosity functions and global 21-cm signals for all HMF/SPS combinations and find that (i) doing so requires additional free parameters to compensate for modeling systematics and (ii) the spread in constraints on parameters of interest for different HMF and SPS choices, assuming $5$ mK noise in the global signal, is comparable to those obtained when adopting the ``true'' HMF and SPS with $\gtrsim 20$ mK errors. Our work highlights the need for dedicated efforts to reduce modeling uncertainties in order to enable precision inference with future datasets.
\end{abstract}
\begin{keywords}
galaxies: high-redshift -- intergalactic medium -- galaxies: luminosity function, mass function -- dark ages, reionization, first stars -- diffuse radiation.
\end{keywords}

\section{Introduction} \label{sec:intro}
Many ongoing and near-future experiments are designed in large-part to provide astrophysical and cosmological constraints on the ``cosmic dawn,'' when the first stars and galaxies formed. For example, galaxy surveys searching for high redshift $z \gtrsim 6$ galaxies will continue piecing together the luminosity function of galaxies over cosmic time \citep{Williams2018,Behroozi2020}, while cosmic microwave background (CMB) measurements constrain the rough timing and duration of reionization from the Thomson scattering optical depth, $\tau_e$, and kinetic Sunyaev-Z'eldovich effect \citep{Planck2018,Reichardt2020}. 21-cm experiments  target neutral hydrogen itself, and are thus direct probes of the mean ionization and thermal histories, as well as their topology \citep{Furlanetto2006,Pritchard2012}.

Of course, the native form of measurements provided by the aforementioned probes, e.g., luminosity functions, $\tau_e$, and 21-cm spectra, are not straightforward to interpret. As a result, forward models of galaxies and reionization are used to predict these quantities, and, once coupled to sampling algorithms like Markov Chain Monte Carlo (MCMC), can be used to infer the underlying parameters of the model from measurements. For example, UV luminosity functions (UVLFs) of high-$z$ galaxies can constrain the star formation efficiency (SFE; $f_{\ast}$) of galaxies \citep{Mason2015,Sun2016}, while the reionization and thermal histories are sensitive to, e.g., the product of $f_{\ast}$, the escape fraction of ionizing photons $\fesc$, and assumptions about galaxies beyond detection limits \citep{Robertson2015,Bouwens2015b}. Meanwhile, 21-cm probes are sensitive also to the minimum mass of halos capable of forming stars, $\Mmin$, and the efficiency of X-ray and non-ionizing UV photon production in high-$z$ galaxies \citep{Barkana2005,Furlanetto2006,Pritchard2007,Mesinger2013}. Additionally, 21-cm experiments can in principle be used to constrain fundamental physics, given that warm and interacting dark matter models affect the formation of structure on small scales \citep{Sitwell2014,LopezHonorez2016,Schneider2018,Munoz2020} and modulate the IGM thermal history relative to the predictions of $\Lambda$ cold dark matter (CDM) models \citep{Tashiro2014,Barkana2018,Munoz2018,Boddy2018,Fialkov2018}. Indeed, many experiments are underway, and have provided interesting limits on both the reionization and thermal histories \citep[e.g.,][]{Bowman2018,Monsalve2017,Singh2018}.

Several parameter inference pipelines have emerged in recent years to interpret the impending deluge of constraints on the cosmic dawn. Semi-empirical or semi-analytic models of galaxy formation are now commonly employed to connect constraints on UVLFs, the mean reionization history, and ionizing background after reionization \citep{Finkelstein2019,Tacchella2018,Behroozi2019,Yung2019a}, as well as global 21-cm signals \citep{Mirocha2017}. Semi-numerical models of reionization have emerged to efficiently model the 21-cm background in 3-D \citep{Mesinger2007,Mesinger2011,Santos2010,Fialkov2013,Hutter2020}, and have recently been coupled to MCMC samplers \citep{Greig2015,Greig2017} and emulators \citep{Kern2017,Cohen2020} to enable joint galaxy survey--21-cm inference \citep{Park2019,Qin2020}.

Recent forecasts largely find that precision constraints on parameters of interest are well within the reach of current facilities. For example, 21-cm measurements can constrain the population-averaged values of $\fstar$, $\fesc$, above $\Mmin$ \citep{Mirocha2015,Greig2017}, warm dark matter \citep{Munoz2020}, and even cosmological parameters \citep{Liu2016a,Kern2017}. If limited only by thermal noise, current experiments are expected to reach the cosmic variance floor in relatively short order \citep{Munoz2021}. Of course, such forecasts are based on ideal instruments, foregrounds, and noise---luxuries yet to be achieved in real experiments, for which systematic uncertainties are the main impediment to progress (see, e.g., \citealt{LiuShaw2020} for a description of some of the main issues).

Systematic uncertainties are not limited to the realm of experiment. Theoretical models are also uncertain, in part due to numerical approximations made to accelerate computations, but also because of assumptions made in the modeling. Some of these assumptions are not necessarily parameterizable, and thus may be difficult to marginalize over when computing parameter constraints. For example, the semi-numerical approach to reionization is an approximation employed to avoid running expensive radiative transfer simulations, and appears to work at the level of $\sim 10-20$\% \citep{Zahn2011,Majumdar2014,Hutter2018}. It is not clear that the level of uncertainty is invariant with respect to model parameters or implementations, and so it is not clear how to generically inflate uncertainties on model parameters of interest. Other aspects of cosmic dawn modeling may be even more difficult to pin down as they are unrelated to numerical approximations, e.g., choosing the ``right'' stellar population synthesis model or halo mass function.

There is no doubt that observational issues are still the most pressing. However, given that the sensitivity of the current generation of 21-cm experiments is sufficient for detection of the global 21-cm signal and power spectrum, a breakthrough in the reduction of systematics could lead to rapid transition from crude upper limits to proper constraints on astrophysical and cosmological parameters. As a result, it is fruitful to consider the limitations of current models, so that particularly weak areas of models can be identified and targeted for priority work. We also hope that doing so may also inspire the community to critically examine the degree of precision needed for different science questions.

While there are many potential sources of systematic uncertainty in models, in this work we focus on three main effects associated with seemingly innocuous modeling decisions: (i) the neglect of uncertainties in cosmological parameters, (ii) the halo mass function (HMF), and (iii) stellar population synthesis (SPS) codes. While cosmological parameter variations can in principle be included self-consistently and marginalized over, they rarely are, especially in semi-numeric codes for which re-generating the cosmology-dependent density and velocity fields can be the most expensive part of the model. Unlike the cosmological parameters, the choice of HMF and SPS model does not lie along a continuum of possibilities. In other words, these ``parameters'' are switches rather than knobs. Other likely causes of uncertainty include numerical techniques (e.g., two-zone vs. semi-numeric vs. fully-numerical), or choice of parameterization for galaxy properties (e.g., mass-dependent vs. mass-independent parameters, inclusion of multiple sources populations). These areas are perhaps equally deserving of attention, though we defer them to future work.

The structure of the paper is as follows. In \S\ref{sec:methods} we review our modeling approach, giving special attention to the components most likely to be uncertain. We present our main results in \S\ref{sec:results} and conclude in \S\ref{sec:conclusions}.

\section{Methods} \label{sec:methods}
Our approach to modeling the reionization and re-heating of the IGM closely follows early two-zone models for the IGM \citep{Furlanetto2006,Pritchard2010a} though we use updated models for the X-ray background and galaxies following \citet{Mirocha2014} and \citet{Mirocha2017}, respectively. We provide a brief summary of the model below, drawing attention to the components sucseptible to the modeling uncertainties we investigate. Our results can be re-created using the publicly-available \textsc{ares} code\footnote{\url{https://ares.readthedocs.io/en/latest/}}.

To begin, we assume that galaxies inhabit dark matter halos in a 1:1 fashion and that star formation is fueld by the inflow of pristine gas from the IGM. This means that the star formation rate $\dot{M}_{\ast}$ (SFR) in galaxies is directly related to the mass accretion rate $\dot{M}_h$ (MAR) of dark matter halos,
\begin{equation}
  \dot{M}_{\ast} = f_{\ast} f_b \dot{M}_h \label{eq:SFR}
\end{equation}
were $f_{\ast}$ is the star formation efficiency (SFE) and $f_b$ is the cosmic baryon fraction. We assume that the SFE as a double-power law (DPL) in halo mass,
\begin{equation}
    \fstar(M_h) = \frac{f_{\ast,10} \ \mathcal{C}_{10}} {\left(\frac{M_h}{M_{\mathrm{p}}} \right)^{-\alphalo} + \left(\frac{M_h}{M_{\mathrm{p}}} \right)^{-\alphahi}} \label{eq:sfe_dpl}
\end{equation}
where $f_{\ast,10}$ is the SFE at $10^{10} M_{\odot}$, $M_p$ is the mass at which $\fstar$ peaks, and $\alphahi$ and $\alphalo$ describe the power-law index at masses above and below the peak, respectively. The additional constant $\mathcal{C}_{10} \equiv (10^{10} / M_p)^{-\alphalo} + (10^{10} / M_p)^{-\alphahi}$ is introduced to re-normalize the standard DPL formula to $10^{10} M_{\odot}$, rather than the peak mass. We model the MAR following \citet{Furlanetto2017}, and assume halos grow at fixed number density, in which case $\dot{M}_h$ can be inferred from the evolution in the HMF. Though idealized, it has the advantage of preserving self-consistency, and does appear to agree well with the results of N-body simulations \citep{Trac2015}, at least for relatively massive halos at $z \lesssim 10$. The MAR is yet another potential source of modeling uncertainty that may be comparable to the effects we explore here \citep{Schneider2020}, though we defer investigation of this possibility to future work.

In this work, we only consider the rest-ultrioviolet continuum emission of galaxies ($1600 \angstrom$), which is probed by recent observations with \textit{Hubble} \citep{Bouwens2015,Finkelstein2015} targeting galaxies at $z \gtrsim 4$. Because the rest-UV continuum probes young stars, it is a good tracer of recent star formation and thus the SFR of galaxies. Under the assumption of a constant SFR, the UV luminosity of a galaxy will asymptote to a constant value on timescales of order $\sim 100$ Myr. In this limit, the UV luminosity can be modeled simply as
\begin{equation}
  \LUV = l_{\mathrm{UV}} \dot{M}_{\ast} \label{eq:LUV}
\end{equation}
where $l_{\mathrm{UV}}$ carries units of $\lsfrunits$ and is often written instead via its reciprocal $\kappa_{\mathrm{UV}} \equiv l_{\mathrm{UV}}^{-1}$.

Finally, with these assumptions, the UV luminosity function is simply
\begin{equation}
  d \phi(\LUV) = \frac{dn_h}{dM_h} \frac{dM_h}{d\LUV} d\LUV . \label{eq:UVLF}
\end{equation}
Here, $dn_h/dM_h$ is the halo mass function, which we compute using the \textsc{hmf} code \citep{Murray2013}, which relies on \textsc{camb} for the matter power spectrum \citep{Lewis2000}. For an assumed value of $l_{\mathrm{UV}}$, one can calibrate $f_{\ast}$ empirically by fitting UVLF constraints from any number of studies \citep{Bouwens2015,Finkelstein2015} using Eqs. \eqref{eq:SFR}-\eqref{eq:LUV} to compute $dM_h/d\LUV$. We note that several aspects of this model have since been generalized, e.g., self-consistent dust reddening, ``bursty star'' formation histories \citep{Mirocha2020a,Mirocha2020b}, though we neglect these complications for the duration of this paper, as they will not obviously exacerbate (or mitigate) the sources of uncertainty we explore here. For results presented in the remainder of this paper, we use models that neglect dust reddening (Fig. \ref{fig:uvlfs}-\ref{fig:rei}; mostly for illustrative purposes), as well as models that employ a relatively standard dust correction using the \citet{Meurer1999} IRX-$\beta$ relation and $\MUV$-$\beta$ relations from \citet{Bouwens2014} (Fig. \ref{fig:params}-\ref{fig:nuisance_pars}). In each case, fits to the relevant datasets are performed using \textsc{emcee} \citep{ForemanMackey2013}.

With a model for sources in hand, we can proceed to modeling the mean reionization history and global 21-cm signal. The differential brightness temperature $\dTb$ is given by
\begin{equation}
    \dTb \simeq 27 (1 - \xHII) \left(\frac{\Obnow h^2}{0.023} \right) \left(\frac{0.15}{\Omnow h^2} \frac{1 + z}{10} \right)^{1/2} \left(1 - \frac{\Tcmb}{\Ts} \right) , \label{eq:dTb}
\end{equation}
where $\xHII = \QHII + (1 - \QHII) x_e$ is the volume-averaged ionized fraction, and
\begin{equation}
    \Ts^{-1} \approx \frac{\Tcmb^{-1} + x_c \Tk^{-1} + x_{\alpha} T_{\alpha}^{-1}}{1 + x_c + x_{\alpha}} . \label{eq:Ts}
\end{equation}
is the spin temperature. The spin temperature quantifies the level populations in the hyperfine singlet and triplet states, and is set by a combination of collisional coupling \citep[quantified by $x_c$;][]{Zygelman2005}, radiative coupling through the Wouthuysen-Field effect \citep[quantified by $x_{\alpha}$;][]{Wouthuysen1952,Field1958} and microwave background temperature, $\Tcmb$. Equation \eqref{eq:dTb} assumes the cosmic mean density and neglects relative velocities along the line of sight, as is appropriate for the global 21-cm signal. 

Our model partitions the IGM into two distinct phases: an ionized phase, characterized by its volume-filling factor $\QHII$, and a ``bulk IGM'' phase, characterized by its temperature and Ly-$\alpha$ intensity. To evolve the gas temperature in the bulk IGM phase, we solve the uniform background problem as outlined in \citet{Mirocha2014}, essentially a high-$z$ implementation of the \citet{Haardt1996} algorithm. Assuming a multi-colour disk spectrum for X-ray emission \citep{Mitsuda1984} with no host-galaxy absorption, we solve for the mean intensity of the cosmic X-ray background as a function of redshift and photon energy, $J_{\nu}(z)$, and integrate over $J_{\nu}$ (weighted by bound-free absorption cross-sections) to obtain the photo-heating and ionization rates as a function of redshift. We assume a fully-neutral IGM, in which case the bound-free optical depth of the IGM to X-ray photons can be tabulated ahead of time for an efficiency boost. The intensity of the Ly-$\alpha$ background, $J_{\alpha}$, which sets the strength of Wouthuysen-Field coupling, is computed in an analogous fashion. Note that this part of the model is cosmology-dependent, as $J_{\nu}$ depends on the Hubble parameter---we self-consistently include the cosmology dependence in this computation. See \S2.2-2.3 in \citet{Mirocha2014} for more details.

Because the mean-free path of ultraviolet photons is short, we relate the growth rate of the fractional ionized volume, $\QHII$, directly to the ionizing photon production rate, i.e.,
\begin{equation}
  \frac{d\QHII}{dt} = \frac{\fesc \Nion \sfrd}{\nH} - \alpha_{\HII} C n_e \QHII
\end{equation}
where $\fesc$ is the escape fraction of ionizing photons, $\Nion$ is the number of ionizing photons emitted per unit stellar mass, and $\sfrd \equiv \int dM_h \sfr dn_h/dM_h$ is the cosmic star formation rate density. The recombination rate is the product of the clumping factor $C$, which we set to unity for simplicity, the electron number density $n_e$, and  the case A recombination coefficient, $\alpha_{\HII}$.

\subsection{Sources of Modelling Uncertainty}
Having established our basic underlying model of UVLFs and the global $21\,\textrm{cm}$ signal, we now highlight several potential sources of theoretical uncertainty.

Latent in virtually every expression in this section are assumptions about cosmological parameters. The 21-cm background itself depends directly on $\Omega_{b,0}$ and the Hubble parameter, as it involves an integral along the line of sight, but also indirectly, as the opacity of the IGM to X-rays similarly depends on $\Omega_{b,0}$, the primordial helium abundance, and the Hubble parameter.

The HMF, while possible to compute analytically under idealized circumstances, is likely more complicated in reality. Recent numerical simulations \citep[e.g.,][]{Tinker2010} suggest that the HMF lies somewhere between the classic analytic models \citep{PressSchechter1974,ShethMoTormen2001} often used in reionization models. The results of N-body simulations can be reproduced via fitting functions, though these vary from model to model due to differences in gravity solver and/or halo finder used in the analysis \citep{Knebe2013a,Knebe2013b}, and themselves may be insufficient for precision work \citep{Bhattacharya2011}. Most numerical simulations run to date are not optimized for high-$z$ \citep[though see, e.g.,][]{Reed2007}, nor has there been a systematic exploration of the agreement between different fitting functions at $z \gtrsim 2$. The fact that variations from one HMF fitting function to the next are larger than those caused by cosmological parameter variations for a single HMF at low-$z$ \citep{Murray2013b} suggests this is a problem worth investigating at high-$z$ as well \citep[see, e.g.,][]{LopezHonorez2016}. In this work, we investigate the \citet{PressSchechter1974}, \citet{ShethMoTormen2001}, and the \citet{Tinker2010} form of the mass function, the latter of which is a representative example that lies between. Other recent results are comparable to \citet{Tinker2010} \citep{Warren2006,Reed2007,Bhattacharya2011}, at least at low redshift where comparisons are most common \citep{Murray2013b}.

Next, the conversion factor from SFR to luminosity, $\kappaUV$ (see Eq. \ref{eq:LUV}), relies on assumptions about stellar evolution, atmospheres, and initial mass function (IMF). Like the HMF, pieces of $\kappaUV$ are in principle ``knobs'' just like the cosmological parameters, e.g., the stellar initial mass function (IMF), which is generally taken to be a power-law or broken power-law. However, in practice, it is common to choose an SPS code and implicitly adopt the IMF used within it. In this work we use the \textsc{bpass} version 1.0 models\footnote{We find that the \textsc{bpass} version 2.1 models \citep{Eldridge2017} differ from the version 1.0 models by $\lesssim 20$\% in the production of UV photons relevant for UVLFs, reionization, and Lyman-$\alpha$ coupling of the 21-cm line for stellar ages $\lesssim 100$ Myr.} \citep{Eldridge2009} as well as the original \textsc{starburst99} models \citep{Leitherer1999}.

\begin{figure}
\begin{center}
\includegraphics[width=0.49\textwidth]{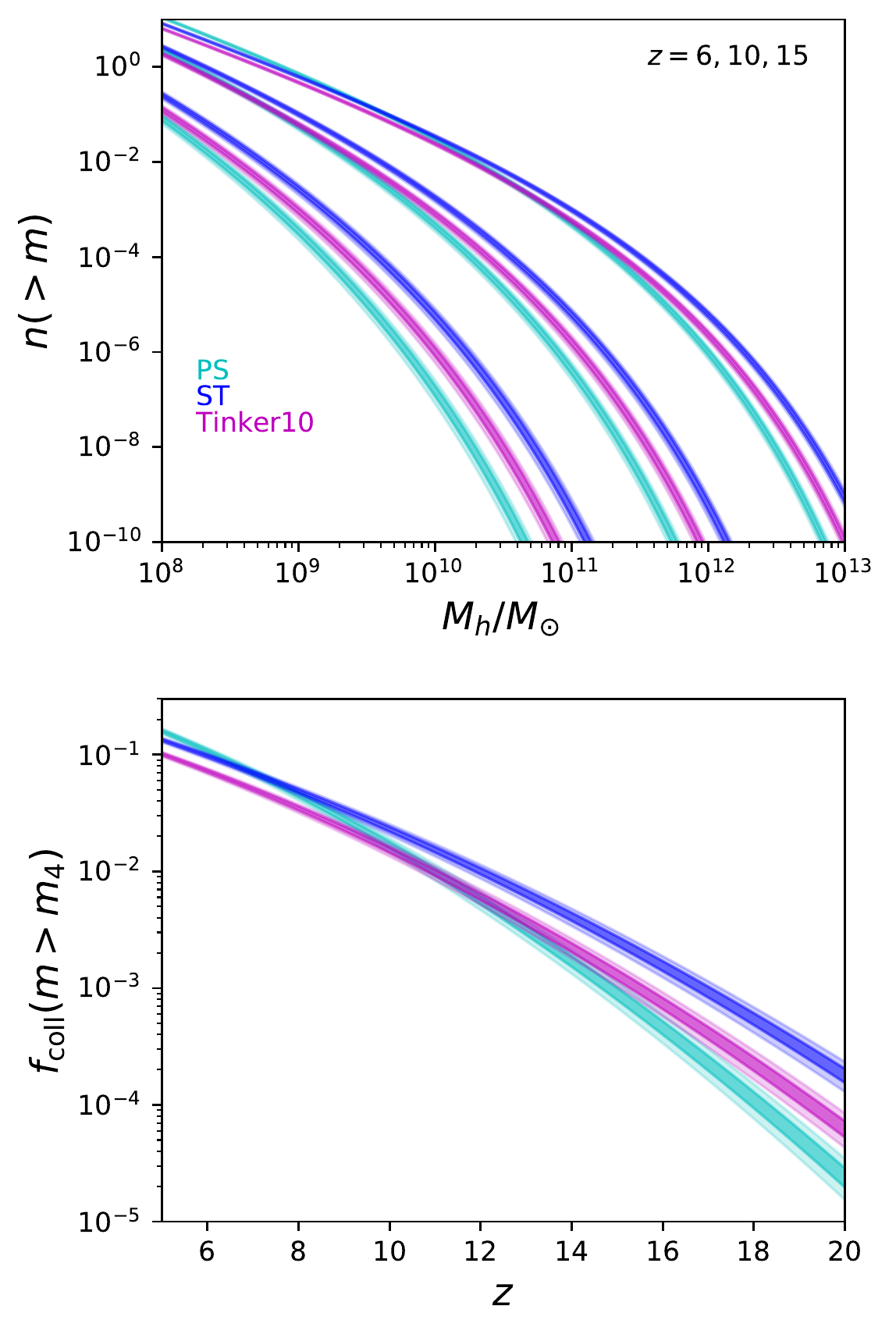}
\caption{{\bf Difference in cumulative HMF (top) and collapse fraction (bottom) for different fitting functions.} Width of each line is due to variations in cosmological parameters ($1$ and $2\sigma$ contours drawn for each), while each colour indicates a different fitting function: \citet{PressSchechter1974} (PS; cyan), \citet{ShethMoTormen2001} (ST; blue), and \citet{Tinker2010} (Tinker10; magenta). In the top panel, the three sets of curves indicate redshifts $z=6,10,$ and 15 from top to bottom.}
\label{fig:hmf}
\end{center}
\end{figure}

In Figure \ref{fig:hmf} we show the HMFs adopted in this work. We show both the cumulative HMF (top) and fraction of matter in collapsed halos, $\fcoll$, above the atomic cooling threshold [$M_h(\Tvir=10^4 \ \mathrm{K}) \equiv m_4$; bottom]. At $z=6$, the disagreement between models is most clear in massive halos, $M_h \gtrsim 10^{11} \ M_{\odot}$, though upon close inspection differences are present in low-mass halos, and the sense of the disparity between fitting functions reverses. At $z \gtrsim 10$, the disagreement is noticeable at all masses. The bottom panel is perhaps the clearer metric for reionization models, where the production rate of UV and X-ray photons is often linked directly to the rate at which mass collapses into halos. At $z \sim 20$, the spread is nearly an order of magnitude, while at $z \sim 6$ it is still a factor of $\sim 2$. In all cases, variations in the cosmological parameters (shaded regions) has a smaller effect than the choice of fitting function (different colours).

\begin{figure}
\begin{center}
\includegraphics[width=0.49\textwidth]{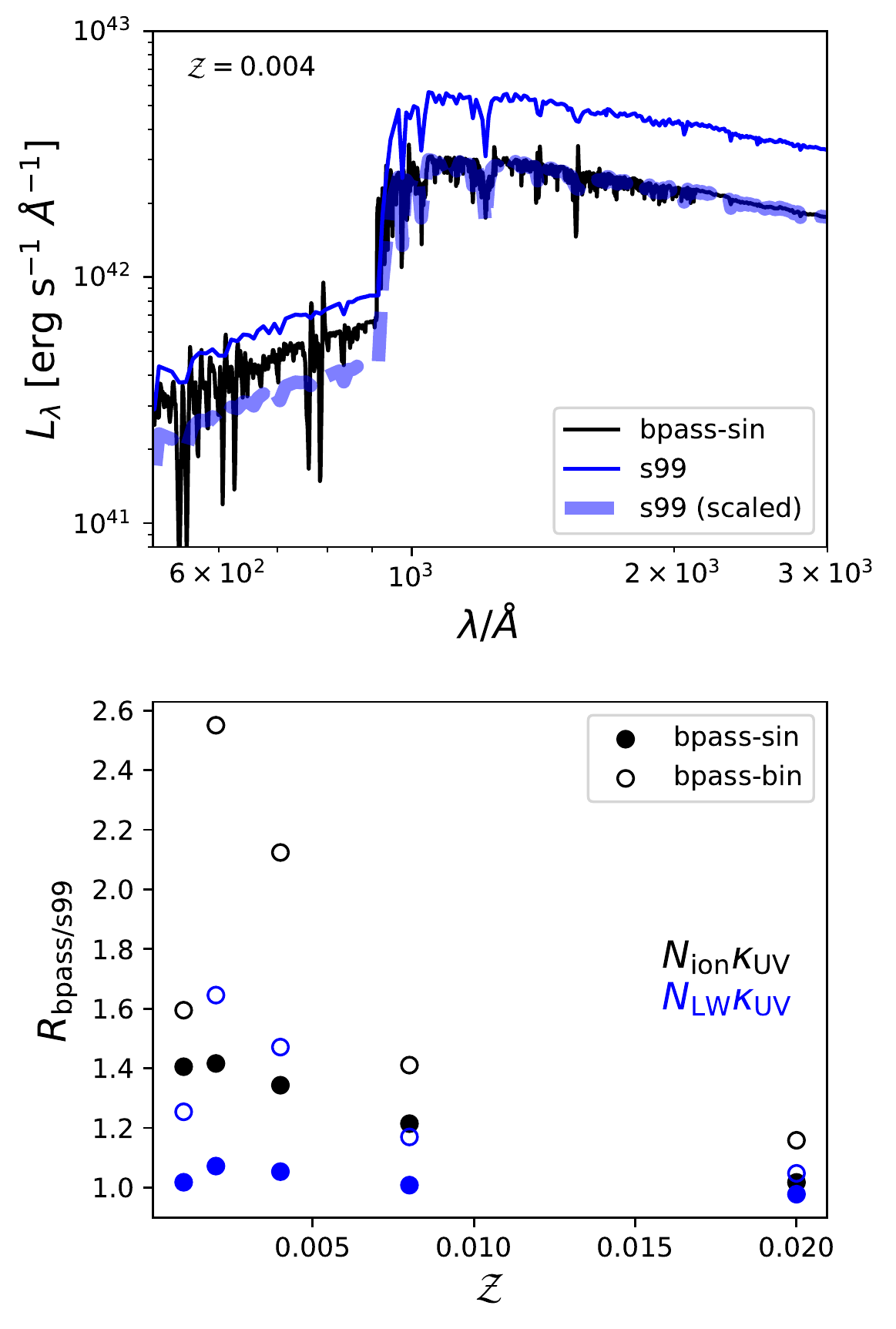}
\caption{{\bf Difference in rest-UV SEDs of galaxies as computed with \textsc{bpass} (black) vs. \textsc{starburst99}.} Top panel shows the SED of a metallicity $\Z=0.004$ stellar population forming stars continuously for the last 100 Myr, while the bottom panel compares the ratio of ionizing ($\Nion$) and Lyman-Werner ($\Nlw$) luminosities to the $1600\angstrom$ continuum for each code at a series of stellar metallicities.}
\label{fig:seds}
\end{center}
\end{figure}

In Figure \ref{fig:seds}, we compare the SPS models used in this work. First, in the top panel, we show an example spectral energy distribution (SED) for a constant SFR and metallicity of $\Z=0.004$ using both \textsc{bpass} (black) and \textsc{starburst99} (blue). There is a clear offset in normalization between the two codes, with \textsc{starburst99} producing more overall emission (per $\sfrunits$) at all wavelengths. To investigate the $\lambda$-dependent nature of the offset, we scale the \textsc{starburst99} spectrum so that it matches \textsc{bpass} at $1600\angstrom$ (blue dashed). It is now clear that, in addition to producing a stronger non-ionizing continuum, \textsc{starburst99} also has a higher \textit{ratio} of non-ionizing to ionizing emission than \textsc{bpass}.

The SED effects are explored further in the bottom panel of Fig. \ref{fig:seds}. For a series of metallicities, we compute the ratio of ionizing (black) and Lyman-Werner band (blue) luminosities to the $1600\angstrom$ luminosity for each code, and then plot the ratio of values for each code, i.e.,
\begin{equation}
  R \equiv \frac{(N \times \kappaUV)_{\mathrm{bpass}}}{(N \times \kappaUV)_{\mathrm{s99}}}
\end{equation}
where $N$ is either $\Nion$, the number of ionizing photons emitted per stellar baryon, or $\Nlw$, the equivalent quantity in the Lyman-Werner band.

The rationale here is that the rest-ultraviolet $\sim 1600\angstrom$ emission is the only part of the spectrum that is (generally) observable---at fixed $\kappaUV$, UVLFs constrain $f_{\ast}$, so any differences in mean reionization history and global 21-cm signal will largely be determined by the product $\Nion \kappaUV$ and $\Nlw \kappaUV$, respectively. Indeed, such differences can be many tens of percent, perhaps a factor of $\sim 2$ in extreme cases, particularly at low metallicity $Z \lesssim 0.01$ and when adopting the \textsc{bpass} binaries models (``bpass-bin''; open symbols). We focus only on single star models throughout this work in order to make a more fair comparison between \textsc{bpass} and \textsc{starburst99}.

\section{Results} \label{sec:results}
We now present our main results, consisting of models run with $10^3$ elements from the \textit{Planck} chain (\texttt{planck\_TTTEEE\_lowl\_lowE}), with three different HMFs, using both \textsc{starburst99} and \textsc{bpass} SPS models. For each cosmological model, we re-generate (i) the recombination history using \textsc{cosmorec} \citep{Chluba2011}, (ii) the optical depth of the IGM used to compute the X-ray background, and (iii) the HMF using \textsc{hmf} \citep{Murray2013}. The cosmological model number is subsequently used as a free parameter, which indicates the appropriate lookup table to use for each of these three quantities, as well as the cosmological parameter values used elsewhere in the model.

\begin{figure}
\begin{center}
\includegraphics[width=0.49\textwidth]{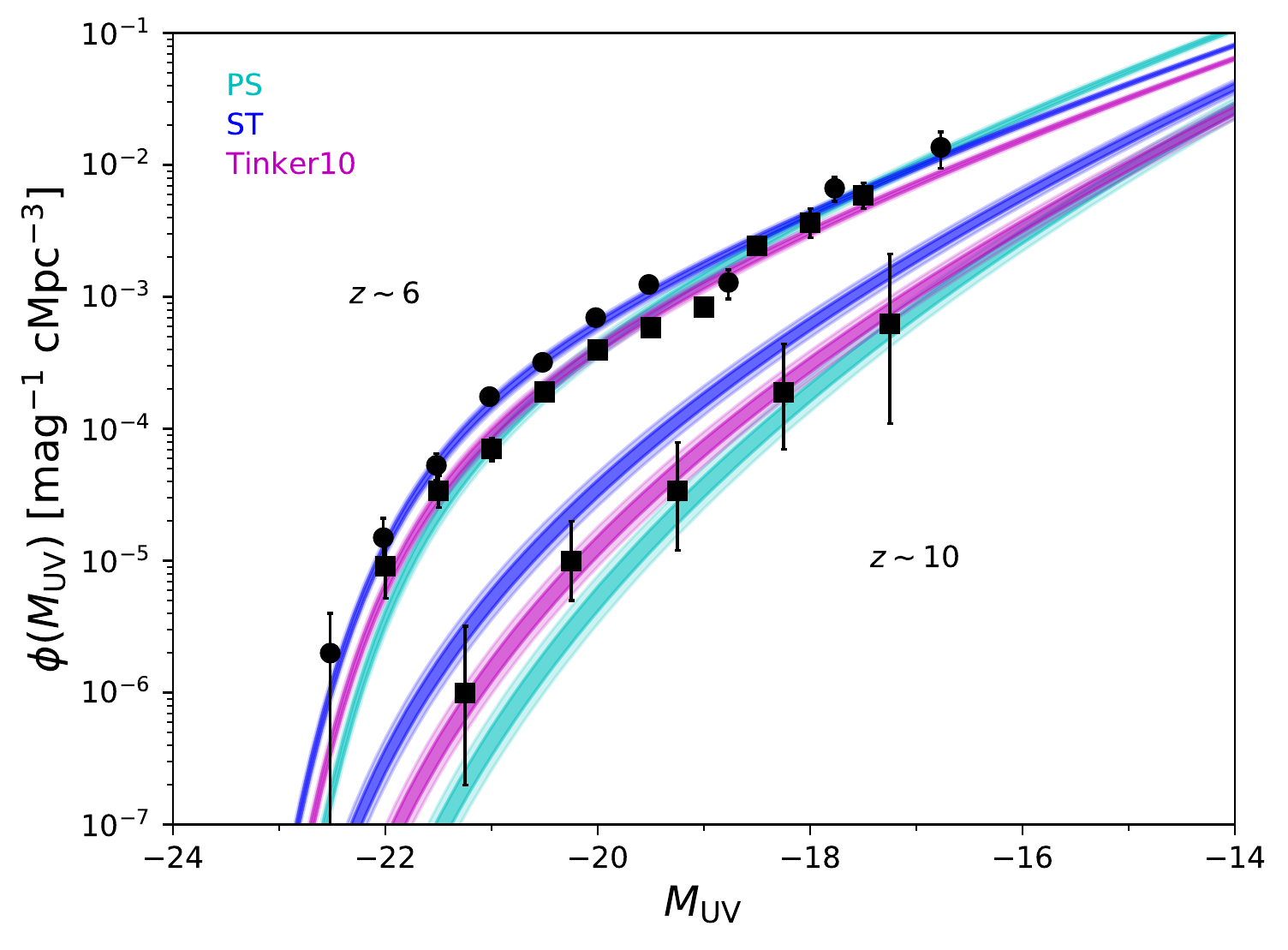}
\caption{{\bf Difference in UVLFs caused by cosmology and HMF variations.} Linestyle and colour conventions are the same as Figures \ref{fig:gs} and \ref{fig:rei}. Data at $z\sim 6$ are from \citet{Bouwens2015} (circles) and \citet{Finkelstein2015} (squares), while $z \sim 10$ measurements are from \citet{Oesch2018}. Note that there are three bands shown for each redshift.}
\label{fig:uvlfs}
\end{center}
\end{figure}

\begin{figure}
\begin{center}
\includegraphics[width=0.49\textwidth]{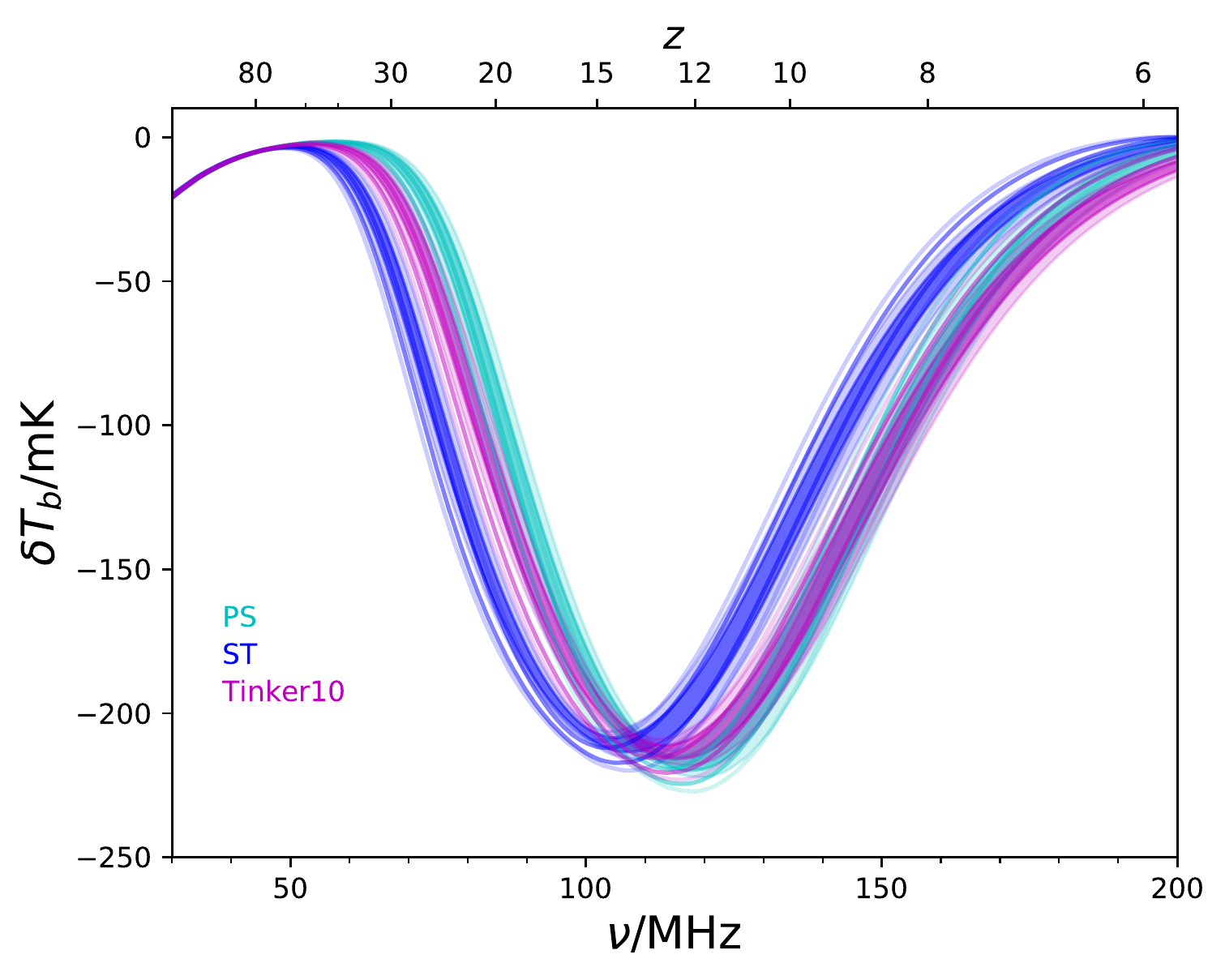}
\caption{{\bf Global 21-cm signal predictions subject to cosmological and HMF uncertainties.} The width of each band is set by uncertainties in the cosmological parameters, while each colour indicates a different HMF fitting function. Open contours use \textsc{bpass} models, while filled contours use \textsc{starburst99}.}
\label{fig:gs}
\end{center}
\end{figure}

\begin{figure}
\begin{center}
\includegraphics[width=0.49\textwidth]{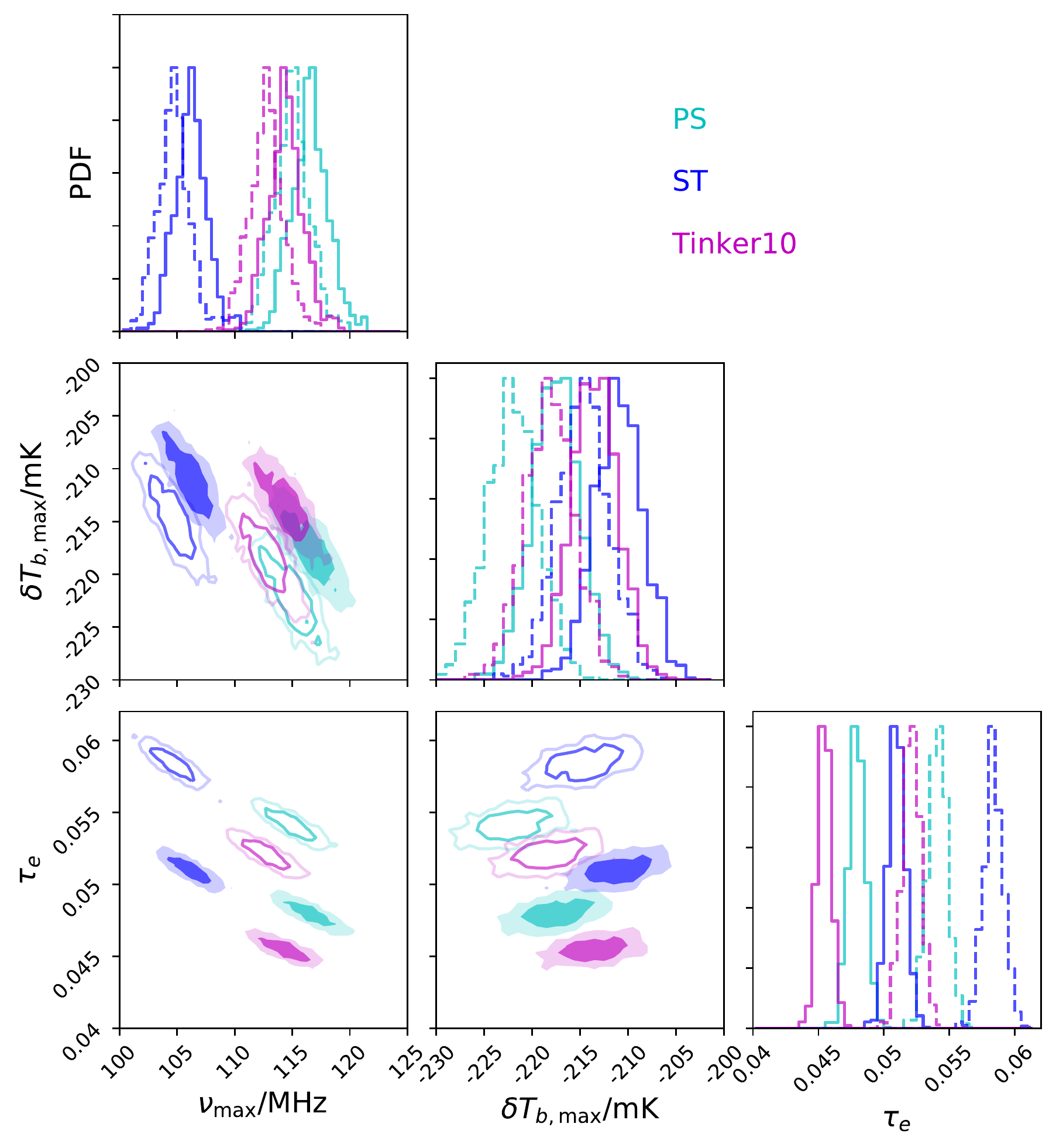}
\caption{{\bf Posterior PDFs of the position of the global 21-cm absorption trough ($\nu_{\min}$, $\delta T_{b,\min})$ and Thomson optical depth.} Different HMF fitting functions are indicated by colour, while open (filled) contours indicate the use of \textsc{bpass} (\textsc{starburst99}). The width of the 68\% and 95\% confidence intervals for each individual HMF/SPS pairing are set by uncertainties in the cosmological parameters.}
\label{fig:triangle}
\end{center}
\end{figure}

\begin{figure}
\begin{center}
\includegraphics[width=0.49\textwidth]{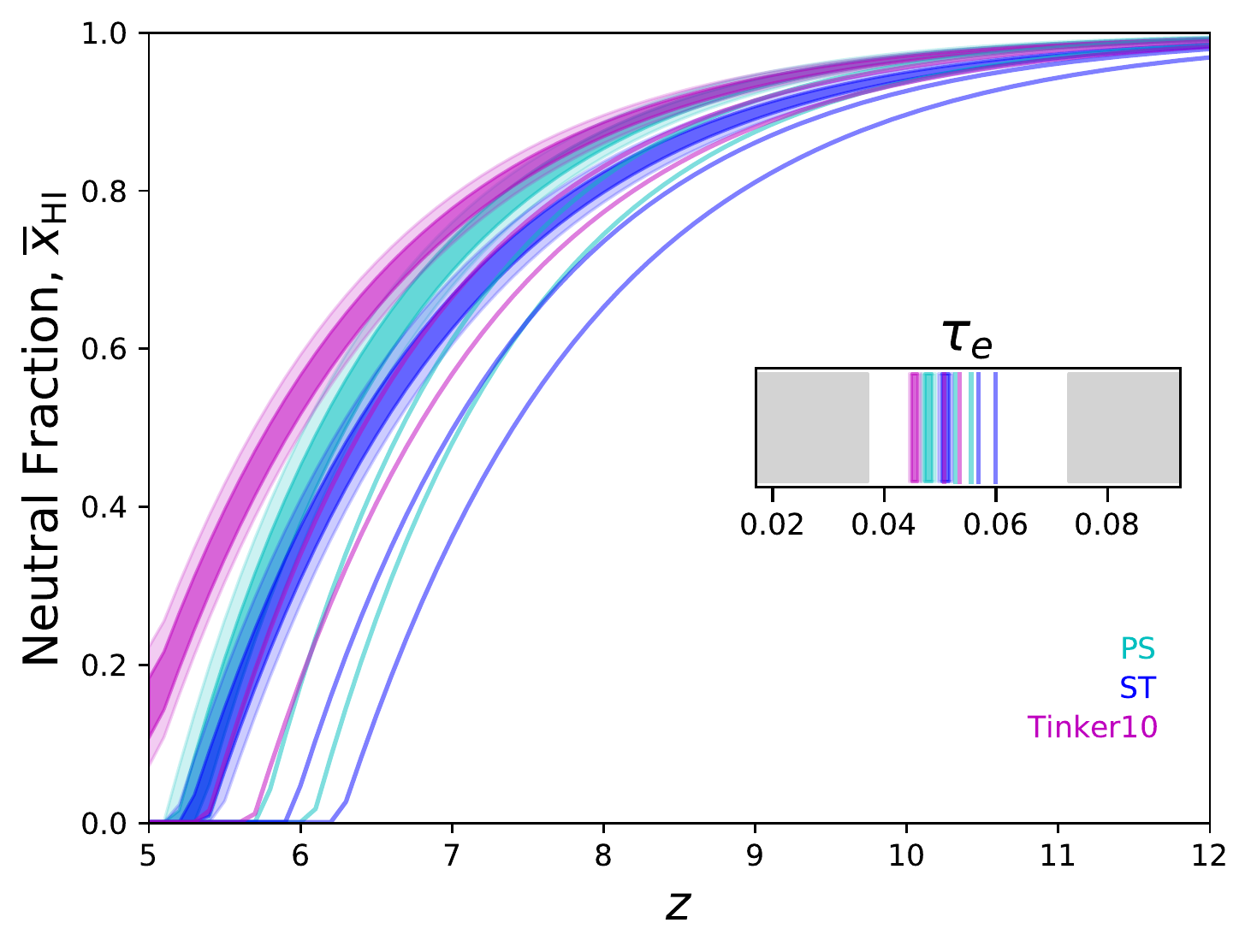}
\caption{{\bf Mean reionization history predictions subject to cosmological and HMF uncertainties.} Conventions are the same as Fig. \ref{fig:gs}. The inset shows the Thomson scattering optical depth shown for each model, with gray regions indicating the range of values excluded at $2\sigma$ according to the \textit{Planck} constraint $\tau_e = 0.055 \pm 0.009$.}
\label{fig:rei}
\end{center}
\end{figure}

\begin{figure*}
\begin{center}
\includegraphics[width=0.98\textwidth]{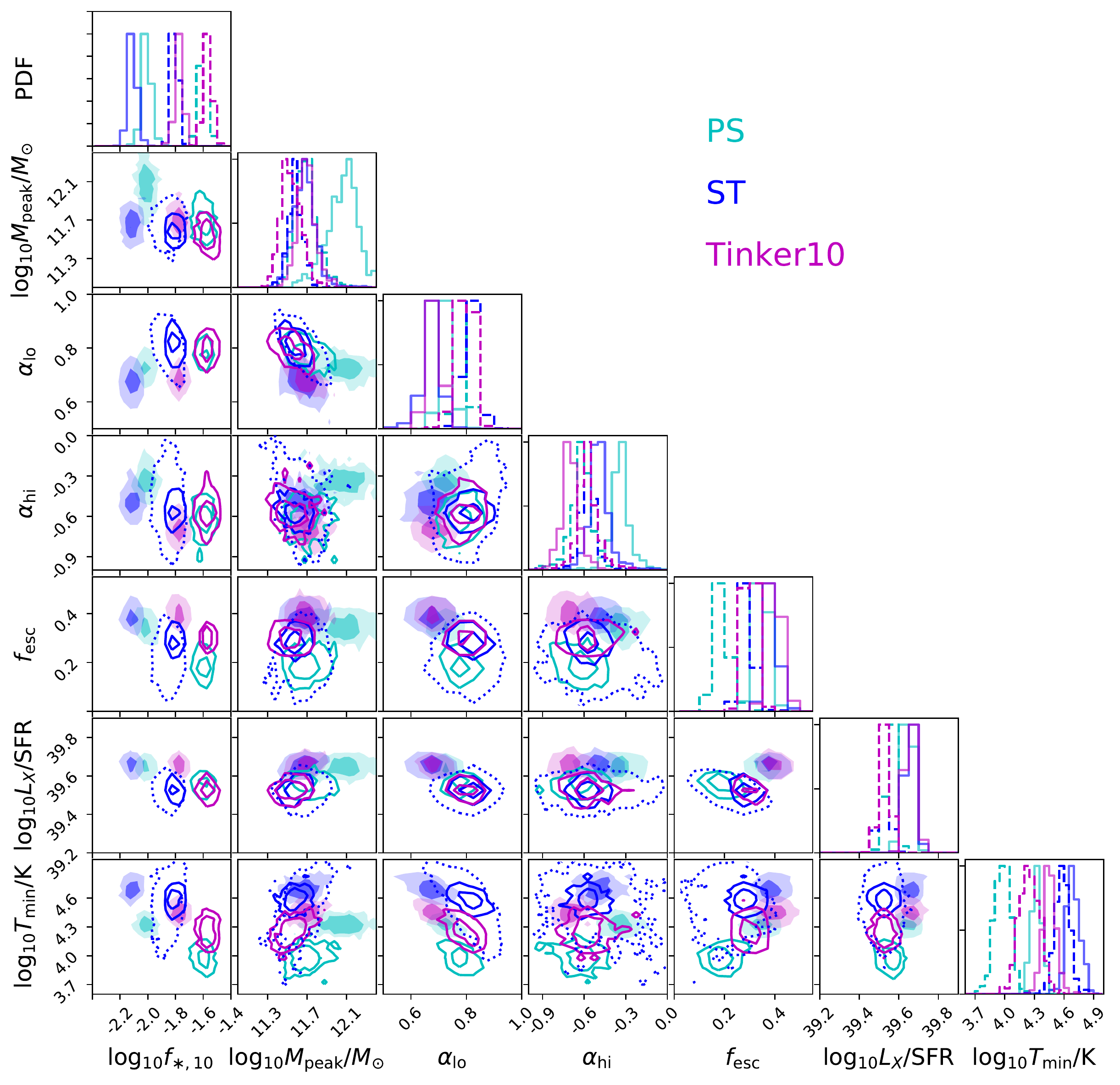}
\caption{{\bf Posterior distributions for six different fits to the same mock global 21-cm signal with $1$ mK Gaussian noise and the $z\sim 6$ UVLF from \citet{Bouwens2015}.} Colours indicate the different HMFs, while filled (open) contours use of \textsc{starburst99} (\textsc{bpass}) SPS models. Marginalized 1-D distributions lie along the diagonal, where dashed lines correspond to the open contours of the interior panels. Dotted contours in the interior panels indicate 95\% contours obtained when assuming 20 mK, rather than 5 mK, Gaussian random noise in the global 21-cm signal mock.}
\label{fig:params}
\end{center}
\end{figure*}

\begin{figure}
\begin{center}
\includegraphics[width=0.49\textwidth]{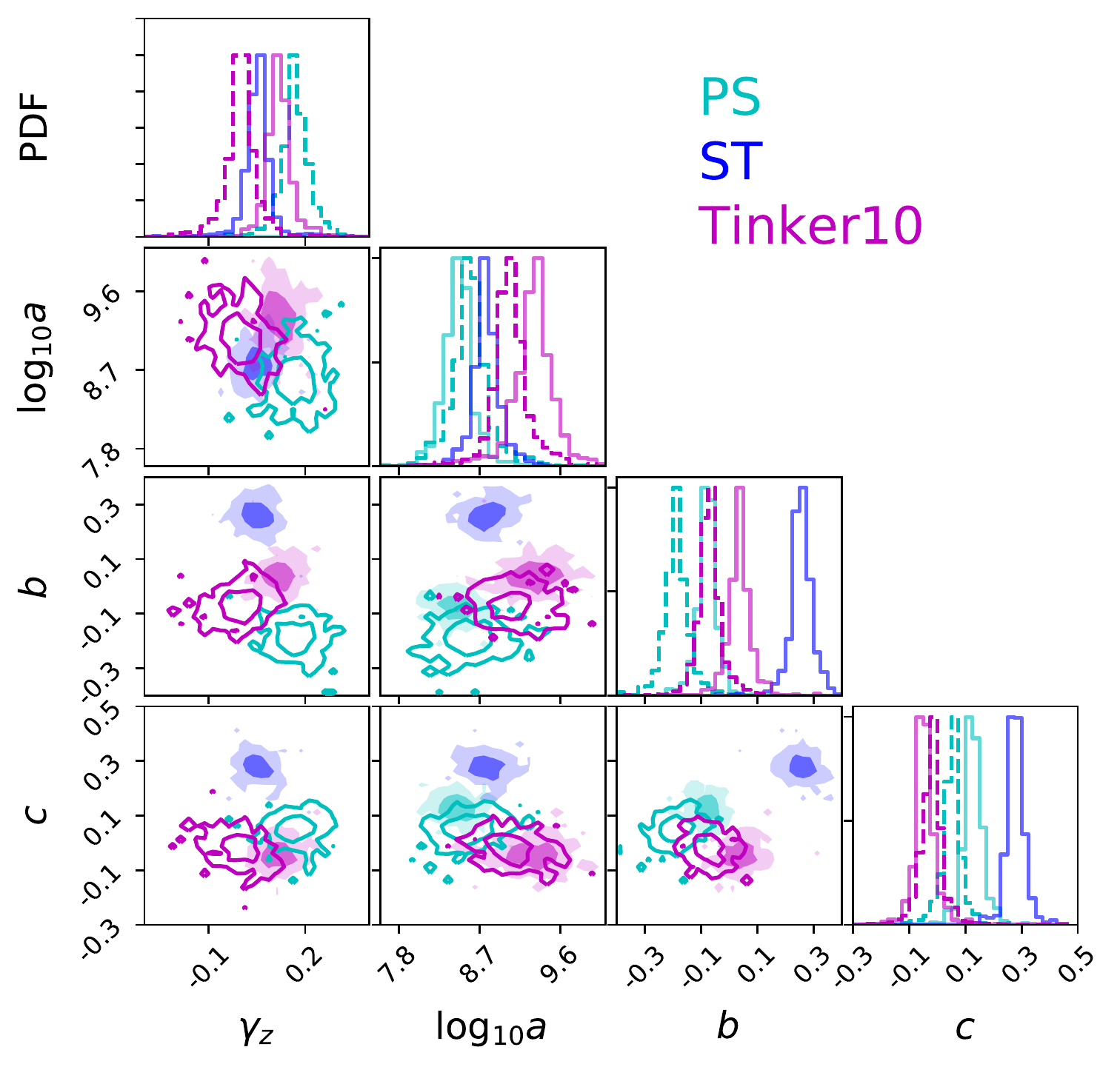}
\caption{{\bf Posterior distributions for four nuisance parameters introduced to relieve tension caused by modeling systematics.} For the mock global 21-cm signal being fit, $f_{\ast}$ does not evolve with redshift ($\gamma_z=0$) and the parameters $a$, $b$, and $c$ (see Eq. \ref{eq:dplx}) are unused.}
\label{fig:nuisance_pars}
\end{center}
\end{figure}

We begin with a pure forward modeling approach to examine the extent to which these different assumptions affect predictions for the mean reionization history, global 21-cm signal, and UVLFs. For this exercise, we adopt the \citet{Mirocha2017} model calibration, which fit the $z \simeq 6$ \citet{Bouwens2015} UVLFs with the model described in \S\ref{sec:methods} under the assumption of no dust reddening (for illustrative purposes). These results are shown in Figures \ref{fig:uvlfs}-\ref{fig:rei}, in which the four parameters governing $f_{\ast}$ (see Eq. \ref{eq:sfe_dpl}) are held fixed to their best-fit values: $f_{\ast,10}=0.05$, $M_{\mathrm{p}} = 2.8 \times 10^{11} \ M_{\odot}$, $\alphalo=0.49$, and $\alphahi=-0.61$. Then, we will turn our attention to the inference problem, and quantify the degree to which parameter constraints are worsened when including these ``new'' sources of uncertainty, while holding roughly fixed the UVLFs and global 21-cm signal via joint fitting. In this case, we employ the standard IRX-$\beta$-based dust correction to more closely mimic the procedure likely to be applied on upcoming measurements. The posterior distribution for each combination of HMF \& SPS model is shown in full in Fig. \ref{fig:params} and \ref{fig:nuisance_pars}.

First, in Figure \ref{fig:uvlfs}, we show the UVLF predictions subjected to cosmological and HMF uncertainties. At $z \sim 6$, the differences between \citet{ShethMoTormen2001} and \citet{Tinker2010} mass functions is comparable to the difference between UVLF measurements reported in \citet{Bouwens2015} and \citet{Finkelstein2015}, though we caution the reader that these two sets of UVLF measurements are obtained via different means and, e.g., make slightly different assumptions about the wavelength corresponding to $\MUV$.

In Figure \ref{fig:gs}, we show predictions for the global 21-cm signal subject to variations in the cosmological parameters and HMF fitting functions. Qualitatively, the curves are very similar---all exhibiting a deep absorption trough at frequencies of $\nu \simeq 110$ MHz, in line with the predictions of \citet{Mirocha2017}. However, the level of variation is not insignificant, as we show more quantiatively in Figure \ref{fig:triangle}. Here, we show only the position of the absorption minimum ($\nu_{\min}$, $\delta T_{b,\min})$, and the Thomson scattering optical depth, $\tau_e$. The position in frequency varies by $\sim 10-15$ MHz between the different fitting functions, while cosmological uncertainties constitute $\sim 2-3$ MHz of these differences for each HMF. Similarly, the amplitude of the absorption trough is uncertain at the level of $\sim 10-15$ mK, about $\sim 5$ mK of which is due solely to uncertainties in cosmological parameters. Finally, $\tau_e$ uncertainties are $\sim 0.01$.

In Figure \ref{fig:rei}, we show the full reionization history for each model. Once again, differences are apparent by eye. For example, at the midpoint of reionization, uncertainties in the cosmological parameters contribute an uncertainty of $\sim 5$\% in the mean neutral fraction, while the position of the reionization midpoint itself varies by $\Delta \zrei \simeq 0.5$ from \citet{ShethMoTormen2001} to \citet{Tinker2010} mass functions. At first glance, the latter seems to produce a reionization history incompatible with current constraints. However, we emphasize that the escape fraction is held fixed in all of these models. Increasing $\fesc$ from 0.2 to $\sim 0.25-0.3$ is enough to bring the \citet{Tinker2010} models into agreement with the \citet{PressSchechter1974} and \citet{ShethMoTormen2001} models. Note that our neglect of dust for this exercise is in part responsible for needing such high escape fractions, since dustier galaxies require more star formation to preserve agreement with UVLFs. We include dust following standard recipes (see \S\ref{sec:methods}) in all that follows.

We now turn our attention to the effect these uncertainties have on our ability to constrain the main parameters of our model. To proceed, we create a global 21-cm signal mock generated assuming the \citet{ShethMoTormen2001} mass function, \textsc{bpass} SPS models, and the best-fit Planck cosmology, with $f_{\ast}$ parameters calibrated to the \citet{Bouwens2015} UVLFs assuming the \citet{Meurer1999} IRX-$\beta$ and \citet{Bouwens2014} $\MUV$-$\beta$ relations to correct for dust. We then add spectrally-flat Gaussian random noise with standard deviation $\sigma=5$ mK to the global 21-cm mock, which is comparable to the level of uncertainty caused by cosmological parameter variations. We adopt the \citet{Bouwens2015} UVLFs throughout as our fiducial set of observations to be fit, though similar conclusions would be drawn if we instead adopted the \citet{Finkelstein2015} measurements. We do not include the $z\sim 10$ UVLFs of \citet{Oesch2018} in the analysis.

There is no guarantee that the global 21-cm signal and, e.g., the $z \sim 6-8$ UVLFs from \citet{Bouwens2015b} can both be well-fit simultaneously if we adopt an HMF or SPS model different from that assumed when generating the 21-cm mock. In order to accommodate this possibility, we employ a simple extension to the standard double power-law $f_{\ast}$ model, multiplying Eq. \eqref{eq:sfe_dpl} by a modulation factor \citep[as in][]{Schneider2020}, i.e.,
\begin{equation}
  f_{\ast} \rightarrow f_{\ast} \bigg[1 + \bigg(\frac{M_h}{a}\bigg)^b \bigg]^c \label{eq:dplx}
\end{equation}
which allows $f_{\ast}$ to depart from a power-law at the low-mass end. This parameterization is appealing because it can capture two qualitatively different possibilities at the low-mass end of the halo population. Depending on the values of $b$ and $c$, the efficiency of star formation can either (i) rapidly decline in halos below $M_h \simeq a$, and thus mimic reionization feedback or other processes which inhibit star formation preferentially in low-mass halos, or (ii) boost $f_{\ast}$ to levels above those predicted by an extrapolation of UVLFs, which could indicate, e.g., very efficient Pop~III star formation and/or a failure of stellar feedback within otherwise normal galaxies.

We further allow $f_{\ast,10}$ to evolve with redshift as a power-law with index $\gamma_z$, resulting in a total of four new parameters. In this work, we treat these as nuisance parameters, though of course departures from power-law and/or redshift-independent $f_{\ast}$ likely correspond to very interesting scenarios worth constraining. Finally, we allow the four usual SFE parameters to vary freely, as well as the normalization of the X-ray luminosity--SFR relation ($L_X/\mathrm{SFR}$), escape fraction of ionizing radiation ($\fesc$), and minimum virial temperature of star-forming halos, $\Tmin$. These last three parameters are only constrained by the global 21-cm signal in our framework, as we do not include measurements of the reionization history in our likelihood.

The results of this exercise are shown in Figures \ref{fig:params} and \ref{fig:nuisance_pars}. First, in each panel of Fig. \ref{fig:params}, we see six posterior distributions: one for each combination of HMF (colours) and SPS model (open vs. filled contours). It is immediately clear that the differences from posterior to posterior are generally larger than the width of any individual posterior distribution. In other words, these systematic modeling uncertainties are larger than the uncertainty caused by the error budget of each individual fit (5 mK Gaussian random noise in global 21-cm signal and errors in UVLF reported by \citealt{Bouwens2015}) as well as uncertainty due to marginalizing over cosmological parameters. For reference, we show also the 95\% contours obtained under the assumption of 20 mK level noise in the global 21-cm signal as dotted contours in each interior panel of Fig. \ref{fig:params}. In some dimensions, the dotted contours enclose a subset of the individual PDFs, indicating that systematic modeling uncertainties are comparable to those caused by a 20 mK noise level. However, generally speaking, the biases causd by modeling systematics are not mimicked well by inflated uncertainties in the global 21-cm signal.

Though Figure \ref{fig:params} paints a somewhat grim picture in which many parameters are constrained much more poorly than idealized forecasts would suggest, a few panels give reasons to be optimistic. For example, the low-mass slope of the SFE (third column), $\alphalo$, while affected by systematics in a relative sense differs little from case to case in an absolute sense---the overall spread is only $\sim 0.1$. For reference, the differences in $\alphalo$ for different stellar feedback scenarios is $1/3$ \citep[see, e.g.,][]{Dayal2013,Furlanetto2017}, while purely empirical models for the SFE vary by $\delta \alphalo \sim 0.5$ depending on assumptions about, e.g., dust \citep{Mirocha2017,Tacchella2018,Mirocha2020a}. As a result, it seems that the part of the SFE most easily related to physical arguments is imminently constrainable.

Similarly, the $L_X$-SFR normalization (sixth column) is also quite impervious to the systematic uncertainties we explored here---though bimodal with respect to SPS model, the difference is order $\sim 0.1$ dex, which is hardly a cause for concern in differentiating models. More worrisome is the minimum virial temperature, $\Tmin$, which varies by an order of magnitude from end to end, and the high-mass behavior of the SFE (set by $\Mpeak$ and $\alphahi$). However, if we can confidently discard the \citet{PressSchechter1974} mass function (cyan contours), the range of viable possibilities in each slice through parameter space is substantially reduced.

Figure \ref{fig:nuisance_pars} focuses on the `nuisance' parameters introduced to alleviate tension caused by modeling systematics. In reality, these parameters are of interest, as they quantify the degree to which the efficiency of star formation departs from a double power-law in mass and whether it may evolve with redshift. Both effects are expected to some degree, as stellar feedback models predict redshift evolution in $f_{\ast}$, while reionization feedback and/or efficient Pop~III star formation are likely to cause a departure from the double power-law form at low-mass.

\section{Discussion \& Conclusions} \label{sec:conclusions}
In this work, we investigated three potential sources of systematic uncertainty in models for high-$z$ galaxies and the mean ionization and thermal history of the IGM.

First, by drawing samples from the \textit{Planck} MCMC chains, we showed that cosmological uncertainties alone result in $\sim 5$\% errors on the mean ionized fraction of the IGM. The global 21-cm signal---which encodes the IGM thermal and ionization histories---is affected by cosmological uncertainties at the level of $\sim 5$ mK in the amplitude of its absorption trough, with $\sim 2-3$ MHz uncertainties in position. These cosmological uncertainties are the smallest of the three effects we explored.

Next, we adopted three commonly-used halo mass function models, \citet{PressSchechter1974}, \citet{ShethMoTormen2001}, and \citet{Tinker2010}, as well as two commonly-used stellar population synthesis models, \textsc{starburst99} \citep{Leitherer1999} and \textsc{bpass} \citep{Eldridge2009}, resulting in a total of 6 possible pairings. Differences due to one's choice of HMF and SPS model are larger than those caused by cosmological uncertainties by virtually any metric. For example, variations in the Thomson scattering optical depth, $\tau_e$, fill roughly half of the $1\sigma$ error bar reported by \textit{Planck}, $\tau_e = 0.055 \pm 0.009$ (Fig. \ref{fig:rei}). The strongest feature of the global signal shifts by up to $\sim 15$ MHz in frequency and $\sim 20$ mK in amplitude (Fig. \ref{fig:gs}).

To explore how these uncertainties propagate to constraints on model parameters, we performed a joint MCMC fit of a mock global 21-cm signal with currenty high-$z$ UVLF constraints. Global 21-cm forecasts based on thermal noise generally predict tight, $\lesssim 10$\% level constraints on model parameters of interest \citep[see, e.g.,][]{Mirocha2015}; however, in many dimensions of parameter space, we find that systematic modeling uncertainties result in factor of $\sim 2$ to $3$ uncertainties in best-fit parameter values. Fortunately, two highly sought-after parameters---$\alphalo$ and $L_X$/SFR---are relatively impervious to systematic uncertainties, leaving the expectations from idealized forecasts largely unchanged.

There are many as-yet-unexplored sources of systematic modeling uncertainty in the reionization context. For example, we have completely neglected any numerical shortcomings of our model, e.g., the accuracy of the two-zone model of the IGM, correlations between the fields composing the 21-cm background, or the idealized approach to high-$z$ galaxy evolution. Of course, more detailed 3-dimensional reionization calculations sourced with galaxies modeled via semi-analytic prescriptions are likely to be more realistic in such regards. However, more sophisticated models have systematic uncertainties as well, caused by, e.g., approximate radiative transfer techniques and sub-grid models for sources and sinks. Similarly, extending models to redshifts $z\ll 6$ can in principle tighten constraints on parameters of interest, but may also introduce yet more systematic uncertainties \citep{Behroozi2010}.

Given the ever-improving constraints on reionization, it is important to continue efforts to quantify the theoretical modeling error budget. Doing so will help to properly report uncertainties on parameters of interest, and identify the biggest problem areas and thus guide dedicated efforts to mitigate them. For some applications, eliminating systematics may be unnecessary, since some parameters of interest cease to be meaningful at arbitrarily high levels of precision. For example, most parameters in 21-cm models are averages over entire galaxy populations, with reference points anchored to uncertain empirical scaling relationships ($L_X$/SFR) for which even constraints at the factor of $\sim 2-3$ level are interesting. However, tests of the cosmological model, e.g., non-cold dark matter, is an intrinsically more quantitative enterprise, in which all sources of systematic uncertainty must be understood or eliminated. In these contexts, there is much work still to be done before limits (or constraints) on various exotic models can be trusted to high precision.

It may be possible to avoid detailed characterization of modeling uncertainties and instead use model selection techniques \citep[e.g., nested sampling;][]{Skilling2006} to constrain modeling assumptions from empirical measurements themselves. There are now several publicly-available tools for model selection \citep[e.g.,][]{Feroz2008,Handley2015}, which are becoming more frequently applied in 21-cm contexts \citep[e.g.,][]{Binnie2019,Sims2020}. While the ultimate goal is to differentiate competing models of galaxy formation and reionization, model selection tools can of course also be used to discern the presence of systematics, including those associated with instruments \citep{Sims2020}, but in principle also systematics associated with signal modeling. The addition of information from 21-cm interferometers would surely help this effort as spatial fluctuations in the 21-cm background are of course also sensitive to the HMF \citep[see, e.g.,][]{LopezHonorez2016}, but also encode the bias of UV and X-ray sources, and should thus mitigate the many degeneracies that impede inference based solely on the mean reionization and thermal histories, like those we focus on here. We defer such analyses to future work.

In this paper we have conducted a first study of how uncertainties in theoretical models can affect interpretations of current and upcoming probes of cosmic dawn. Although the conclusions may seem pessimistic at first glance, they in fact demonstrate the excitement that will accompany the expected observational progress of the next few years: rather than a scenario where experiments will merely fine tune a handful of model parameters, we will be in a situation where there exists a truly symbiotic relationship between theory and experiment, with each side informing the other in future improvements. Together, they will lead the community toward a new understanding of our cosmic dawn.

\section*{Acknowledgments}
The authors thank Saurabh Singh, Brad Greig, Peter Sims, and Elizabeth Stanway for helpful conversations that improved this manuscript, and the anonymous referee for many useful suggestions that helped clarify the text. J.M. acknowledges support from a CITA National Fellowship. H.L. was supported by a Science Undergraduate Research Award from the McGill University Faculty of Science. A.L. acknowledges support from the New Frontiers in Research Fund Exploration grant program, a Natural Sciences and Engineering Research Council of Canada (NSERC) Discovery Grant and a Discovery Launch Supplement, a Fonds de recherche Nature et echnologies Quebec New Academics grant, the Sloan Research Fellowship, the William Dawson Scholarship at McGill, as well as the Canadian Institute for Advanced Research (CIFAR) Azrieli Global Scholars program. Computations were made on the supercomputer Cedar at Simon Fraser University managed by Compute Canada. The operation of this supercomputer is funded by the Canada Foundation for Innovation (CFI).

\textit{Software:} numpy \citep{numpy}, scipy \citep{scipy}, matplotlib \citep{matplotlib}, h5py\footnote{\url{http://www.h5py.org/}}, and mpi4py \citep{mpi4py1}.

\textit{Data Availability:} The data underlying this article is available upon request, but can also be re-generated from scratch using the publicly available \textsc{ares} code.

\bibliography{references}
\bibliographystyle{mn2e_short}

\end{document}